\documentclass[11pt,a4paper]{article}
\usepackage{jinstpub}

\usepackage{graphicx}
\usepackage[title]{appendix}
\usepackage{booktabs}
\usepackage{array}
\usepackage{threeparttable}
\title{Development and characterization of a wingless ICPC HPGe detector with thin amorphous-Ge contacts}

\author[a]{S.~A.~Panamaldeniya,}
\author[a]{K.~M.~Dong,}
\author[a,1]{and D.~M.~Mei%
\note{Corresponding author.}}
\affiliation[a]{Department of Physics, University of South Dakota, 414 E. Clark St., Vermillion, SD 57069, U.S.A.}

\emailAdd{dongming.mei@usd.edu}

\abstract{
High-purity germanium (HPGe) detectors with thin amorphous-germanium (a-Ge) contacts can provide bipolar charge blocking and surface passivation while minimizing contact-related inactive thickness. We report the fabrication and characterization of a p-type inverted coaxial point-contact (ICPC) HPGe detector using thin a-Ge contacts in a wingless geometry that avoids the protective wing structure used in our previous prototype. The detector was fabricated from a USD-grown, zone-refined HPGe crystal and has a mass of 20.0~g. At 77~K, it exhibited picoampere-scale leakage current and an operational depletion voltage of approximately 440~V. Gamma-ray spectroscopy yielded energy resolutions of 1.75~keV FWHM at 59.5~keV and 3.19~keV FWHM at 662~keV. The calibrated charge-injection method gave an absolute capacitance of 0.496~pF, with an estimated method-level relative uncertainty of 6--8\%, consistent with the 0.488~pF value obtained from electrostatic modeling with \texttt{SolidStateDetectors.jl}. With a $^{241}$Am source positioned inside the detector bore, the detector showed a source-correlated excess current and a broad degraded-alpha continuum centered near 3558~keV, demonstrating sensitivity to near-surface interactions. Because the source encapsulation, intervening materials, and near-surface charge-collection response were not independently characterized, the continuum is not used to infer a unique dead-layer thickness or energy-loss mechanism. The principal result is successful fabrication and stable sub-pF operation of a 20-g wingless thin-contact ICPC prototype.
}

\keywords{Inverted coaxial point-contact detectors, high-purity germanium (HPGe), amorphous-Ge contacts, low capacitance, near-surface radiation detection}

\begin{document}
\maketitle
\flushbottom

\section{Introduction}

High-purity germanium (HPGe) detectors are widely used in gamma-ray spectroscopy  ~\cite{Hafizoglu2024}, nuclear structure studies ~\cite{Zakoucky2004}, homeland security ~\cite{Stave2015}, and rare-event physics ~\cite{Mei2006_Muon, Agnese2018_SCDMS} because of their excellent energy resolution and high detection efficiency. Among the various detector geometries, inverted coaxial point-contact (ICPC) detectors combine the large active volumes of coaxial detectors with the low capacitance and low electronic noise of point-contact designs, making them attractive for low-background applications ~\cite{Cooper2011_PSA}.

The performance of HPGe detectors is strongly influenced by the contact technology. Conventional p-type HPGe detectors typically employ a lithium-diffused $n^{+}$ contact and a boron-implanted $p^{+}$ contact ~\cite{Bhatt2014_GeJunction}. While lithium-diffused contacts provide effective charge blocking, they introduce a thick dead layer that reduces the active detector volume and attenuates low-energy radiation ~\cite{Kaya2022_DeadLayer}. Thin amorphous-germanium (a-Ge) contacts provide an attractive alternative, offering bipolar charge blocking and effective surface passivation while limiting contact-related inactive thickness~\cite{Meng2019_aGe, Amman2018_aGe, Looker2014_Thesis}.

In our previous work~\cite{Panamaldeniya2026_ICPC}, we demonstrated an ICPC HPGe detector employing thin a-Ge contacts with a winged support structure to protect the fragile contact surfaces during fabrication and handling. However, the winged design required additional machining of detector-grade germanium. In the present work, we investigate a wingless ICPC detector design to simplify fabrication and avoid wing-related machining while maintaining detector performance. The material-utilization benefit is treated qualitatively because the archived machining records do not permit a reliable reconstruction of the mass removed specifically to form the earlier wing structure.

In addition, the very thin a-Ge contact structure~\cite{Looker2015_Leakage, Quang2011_DeadLayer, Amman2018_aGe, Amman2007_aGeContacts} can substantially reduce contact-related attenuation and thereby enable studies of near-surface alpha-particle interactions that are generally inaccessible through a thick Li-diffused outer contact. We report the fabrication and characterization of a wingless thin-contact ICPC detector and investigate its response to an internal $^{241}$Am source.

Relative to our previous thin-contact ICPC study~\cite{Panamaldeniya2026_ICPC}, the present work provides four specific advances: (i) fabrication and cryogenic operation of an ICPC detector without the protective wing structure; (ii) scaling from the 5-g SAP17 prototype to the 20-g SAP22 detector while retaining sub-pF capacitance; (iii) a quantitative consistency comparison between the measured absolute capacitance and a \texttt{SolidStateDetectors.jl} electrostatic model; and (iv) characterization under localized $^{241}$Am irradiation inside the ICPC bore, including observation of a degraded-alpha continuum. Although SAP22 is substantially larger than SAP17, it remains a prototype-scale ICPC detector; extending the wingless thin-contact process to much larger detector masses is an important future scaling step.

\section{Detector design and fabrication}
\label{sec:DETECTOR DESIGN AND FABRICATION}

\subsection{Detector fabrication}

The detector (SAP22) was fabricated from a USD-grown, zone-refined p-type HPGe crystal using the same surface-preparation, amorphous-Ge contact fabrication, and metallization procedures reported in our previous work \cite{Panamaldeniya2026_ICPC}. Briefly, the crystal was mechanically shaped, polished using Al$_2$O$_3$ slurries and SiC abrasive papers, chemically etched in a HF:HNO$_3$ (1:4) solution to remove surface damage, and subsequently coated with amorphous Ge (a-Ge) by RF sputtering in a 7\% H$_2$/Ar atmosphere. Aluminum contacts were then deposited by DC magnetron sputtering to form the point-contact and outer-electrode regions.

The nominal a-Ge and Al thicknesses were approximately 600~nm and 120~nm, respectively, consistent with Ref.~\cite{Meng2019_aGe}. The point contact was defined using a localized HF etch through a masking template. No post-deposition annealing was performed. Detailed descriptions of the polishing procedure, chemical etching, sputtering conditions, contact definition, and cryogenic characterization methods can be found in Ref.~\cite{Panamaldeniya2026_ICPC}.

The primary difference from Ref.~\cite{Panamaldeniya2026_ICPC} is that the present detector geometry does not incorporate the winged handling structure previously introduced to facilitate detector manipulation during fabrication. The finished wingless detector has a mass of 20.0~g; this mass is included with the geometry parameters in Table~\ref{tab:geometry}.

\subsection{Detector geometry}

The detector employs an inverted coaxial point-contact (ICPC) geometry similar to that reported previously in Ref.~\cite{Panamaldeniya2026_ICPC}. The detector consists of a cylindrical HPGe crystal containing a shallow coaxial bore and a small point contact. Unlike the earlier design, the present detector was fabricated without the winged handling structure. All other contact technologies, surface treatments, and fabrication procedures remained unchanged. Key geometrical and fabrication parameters are summarized in Table~\ref{tab:geometry}.

\begin{table}[htbp]
\centering
\caption{Geometrical and fabrication parameters of the wingless ICPC HPGe detector.}
\label{tab:geometry}
\begin{tabular}{lc}
\hline
\textbf{Parameter} & \textbf{Detector} \\
\hline
Crystal material & p-type HPGe \\
Detector body diameter (mm) & 25.5 \\
Detector height (mm) & 10.0 \\
Detector mass (g) & 20.0 \\
Coaxial bore diameter (mm) & 17.7 \\
Coaxial bore depth (mm) & 4.2 \\
Point contact diameter (mm) & 1.5 \\
Circumferential groove depth (mm) & 3.0 \\
a-Ge coated surfaces & All exposed Ge surfaces \\
Metallized surfaces & Point contact and outer contact \\
Nominal a-Ge thickness (nm) & 600 \\
Nominal Al thickness (nm) & 120 \\
\hline
\end{tabular}
\end{table}

\begin{table}[htbp]
\centering
\caption{Direct comparison of the previously reported winged ICPC detector
SAP17 and the present wingless ICPC detector SAP22. SAP17 values are taken
from Ref.~\cite{Panamaldeniya2026_ICPC}.}
\label{tab:SAP17_SAP22_comparison}
\small
\begin{tabular}{>{\raggedright\arraybackslash}p{0.40\textwidth}
                >{\centering\arraybackslash}p{0.22\textwidth}
                >{\centering\arraybackslash}p{0.27\textwidth}}
\toprule
\textbf{Quantity} &
\textbf{SAP17 (winged)} &
\textbf{SAP22 (wingless)} \\
\midrule
Body diameter (mm) & 14.4 & 25.5 \\
Detector height (mm) & 7.8 & 10.0 \\
Detector mass (g) & 5.0 & 20.0 \\
Spectroscopy operating bias (V) & 400 & 700 \\
Experimental capacitance (pF) & 0.503 & 0.496 \\
Simulated capacitance (pF) & 0.537 & 0.488 \\
Nominal experiment--simulation difference (\%) & 6.76 & 1.61 \\
59.5-keV FWHM (keV) & 2.10 & 1.75 \\
662-keV FWHM (keV) & 4.50 & 3.19 \\
Fabrication/handling outcome & Successful with wings & Successful without wings \\
Alpha-response study & Not reported & Broad degraded continuum near 3558~keV \\
\bottomrule
\end{tabular}
\end{table}

The comparison in Table~\ref{tab:SAP17_SAP22_comparison} highlights the
scaling from the previously reported winged SAP17 detector to the present
wingless SAP22 detector. Despite its substantially larger dimensions and
mass, SAP22 retains sub-pF capacitance: the experimentally extracted value
is 0.496~pF, compared with 0.503~pF for SAP17. The charge-injection
capacitance values carry an estimated relative method uncertainty of
approximately 6--8\%, so the numerical experiment--simulation differences
listed in the table should be regarded as nominal consistency checks rather
than precision validation at the percent level. SAP22 also exhibits narrower
photopeaks than SAP17 at both 59.5 and 662~keV under the source and bias
conditions reported for each device. These results demonstrate that removal
of the protective wing structure and scaling to a larger detector did not
prevent low-capacitance operation or prototype-level spectroscopy. The
comparison should not be interpreted as demonstrating that removal of the
wings itself improves energy resolution, because detector dimensions,
operating bias, and other device-specific factors also differ. In addition,
the best spectroscopic performance in the previous study was obtained with
SAP16; therefore, the principal demonstrated advantage of SAP22 remains
successful fabrication and handling of a substantially larger detector
without wings. The material-utilization advantage is presently qualitative
because the available machining records do not permit a defensible
reconstruction of the mass removed specifically to form the earlier wing
structure.

\begin{figure}
    \centering
    \includegraphics[width=1\linewidth]{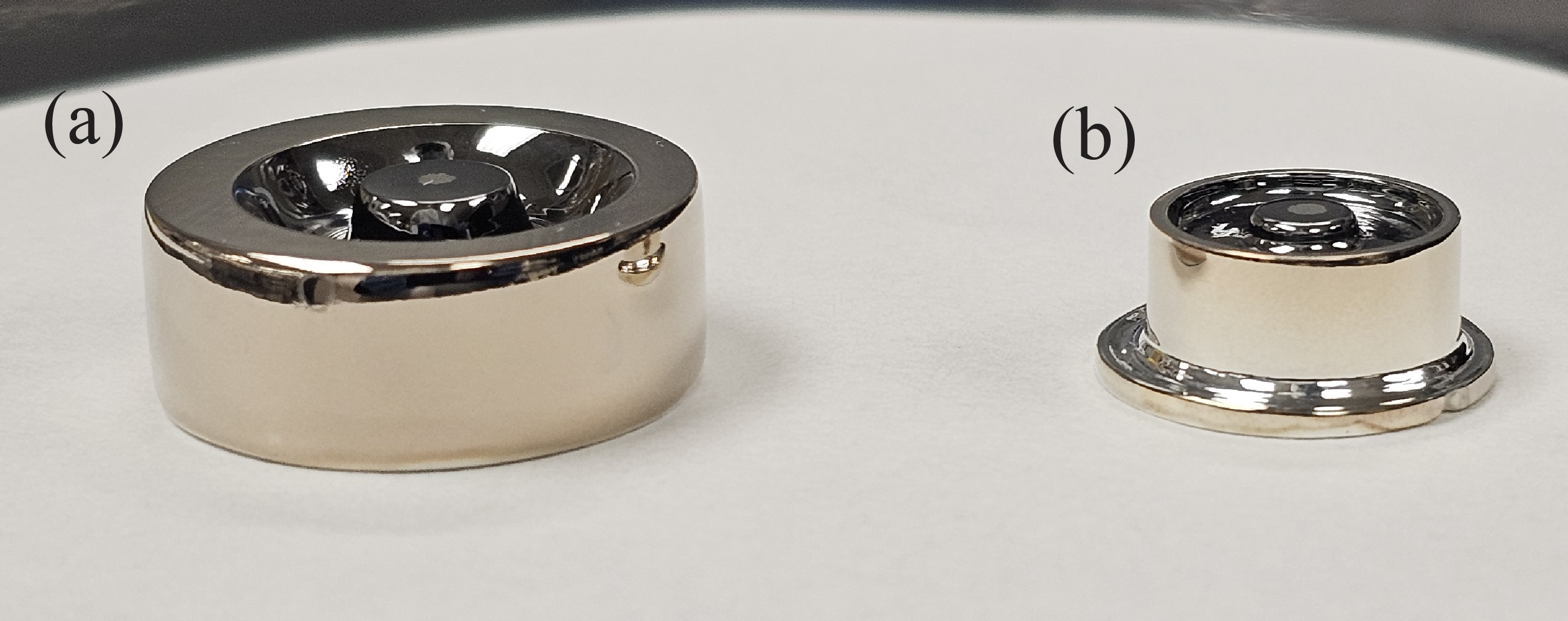}
    \caption{Photographs comparing the ICPC HPGe detector geometries:
(a) the present wingless detector (20.0~g) and (b) the previously developed winged detector (5.0~g) ~\cite{Panamaldeniya2026_ICPC}. The winged structure was originally introduced to facilitate detector handling during fabrication, whereas the present design demonstrates successful fabrication and handling
of a substantially larger detector without these protective structures.}
    \label{fig:placeholder}
\end{figure}

\section{Experimental setup and analysis methods}
\label{sec:Experimental setup and analysis methods}

\begin{figure}
    \centering
    \includegraphics[width=1\linewidth]{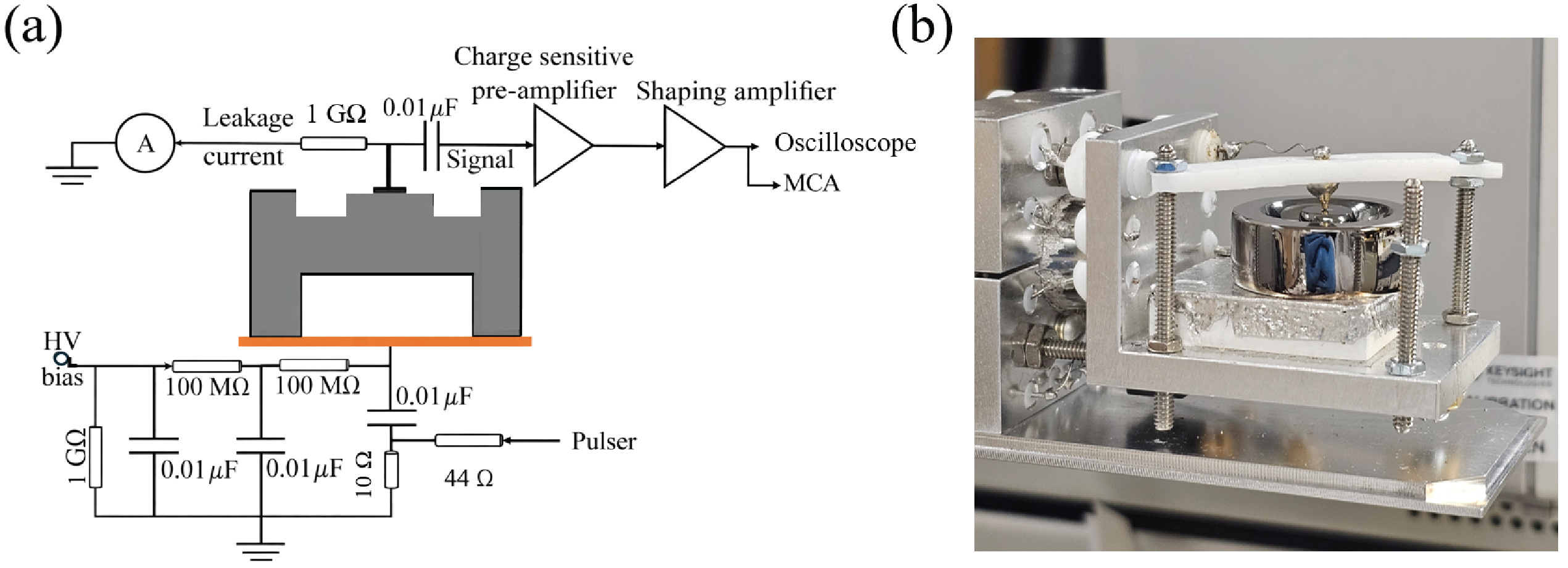}
    \caption{(a) Schematic of the electronics chain used for detector characterization and spectroscopy. (b) Photograph of the detector mounted inside the cryostat. The pulser-injection branch contains 44~$\Omega$ and 10~$\Omega$ resistors and was used for capacitance calibration and signal checks. Source-based measurements were performed through the normal detector readout chain.}
    \label{circuit diagram}
\end{figure}

\subsection{Cryogenic environment and electronics chain}

All measurements were performed in a liquid-nitrogen-cooled cryostat operated at a stabilized temperature of 77~K. A positive high-voltage bias was applied to the outer contact, while the point contact was held at ground potential and used for signal readout.

The cryogenic system, electronics chain, and measurement procedures were identical to those reported in Ref.~\cite{Panamaldeniya2026_ICPC}. Leakage-current measurements were performed using a Keithley 6482 picoammeter, and spectroscopic measurements were acquired using an ORTEC 927 multichannel analyzer. 
Unless otherwise noted, a shaping time of $1~\mu\mathrm{s}$ was used; the capacitance-calibration measurement described in Section~\ref{sec:Capacitance measurement and interpretation} was acquired using a shaping time of $2~\mu\mathrm{s}$. A schematic of the electronics chain and a photograph of the detector mounted inside the cryostat are shown in Figure~\ref{circuit diagram}.

\subsection{Capacitance measurement and interpretation}
\label{sec:Capacitance measurement and interpretation}

The bias-dependent C--V response was measured with the same charge-injection branch and fixed electronics configuration used for the absolute-capacitance calibration. At each bias point, the pulser response was converted to an effective capacitance using the calibrated charge-injection relation described below; for display in Figure~\ref{IV}(b), the resulting values were normalized to the lowest-bias value in the corresponding scan. The normalized trace is therefore a dimensionless relative C--V metric. Because each source-present and source-free scan is normalized independently to its own lowest-bias value, Figure~\ref{IV}(b) is intended for comparison of the bias dependence and plateau behavior rather than for comparison of absolute capacitance between the two source configurations. In the absence of a radioactive source, the relative capacitance decreases with increasing bias and approaches a near-constant plateau at approximately 440~V. Because the plateau onset is gradual in this non-planar geometry and the measurement was sampled at discrete bias values, 440~V is reported as an \emph{operational depletion voltage} rather than as a sharply determined threshold with a formal statistical uncertainty.

The absolute detector capacitance was determined using the calibrated charge-injection method described in Ref.~\cite{Panamaldeniya2026_ICPC}. The pulser amplitude was monitored using a Keysight InfiniiVision DSOX3034A digital oscilloscope, and the MCA energy scale was calibrated using the 662-keV photopeak from $^{137}$Cs. The pulser-injection branch shown in Figure~\ref{circuit diagram} contains 44~$\Omega$ and 10~$\Omega$ resistors, so the voltage delivered to the detector-side injection node is

\begin{equation}
V_{\mathrm{inj}}=V_{\mathrm{pulser}}\frac{10}{44+10}.
\label{eq:vinj}
\end{equation}

For $V_{\mathrm{pulser}} = 50~\mathrm{mV}$, Eq.~\ref{eq:vinj} gives $V_{\mathrm{inj}} = 9.26~\mathrm{mV}$. The capacitance-calibration measurement was acquired using a shaping time of $2~\mu\mathrm{s}$, and the same pulser setting produced a peak at approximately $85~\mathrm{keV}$ on the gamma-ray-calibrated MCA scale. This calibration measurement was performed separately from the spectroscopy measurements presented in Section~\ref{sec:Gamma-ray spectroscopy performance}, for which a shaping time of $1~\mu\mathrm{s}$ was used. Accordingly, the pulser peak position in the capacitance-calibration measurement should not be directly compared with the pulser peak positions shown in Figures~\ref{both outside} and ~\ref{Am241 inside}.

Using the average pair-creation energy in Ge at 77~K, $\varepsilon_{\mathrm{Ge}}=2.96~\mathrm{eV}$, the equivalent injected charge is

\begin{equation}
Q_{\mathrm{inj}}=e\frac{E_{\mathrm{pulser}}}{\varepsilon_{\mathrm{Ge}}}
\simeq 4.60~\mathrm{fC},
\label{eq:qinj}
\end{equation}

and the detector capacitance is

\begin{equation}
C_{\mathrm{det}}=\frac{Q_{\mathrm{inj}}}{V_{\mathrm{inj}}}
\simeq 0.497~\mathrm{pF}.
\label{eq:cdet}
\end{equation}

Using the unrounded calibration values from the analysis yields the reported value of 0.496~pF. The dominant uncertainty sources are the pulser-energy calibration, voltage calibration/resolution, resistor tolerances, and stability of the electronics chain. The same charge-injection procedure was estimated in Ref.~\cite{Panamaldeniya2026_ICPC} to carry a relative uncertainty of approximately 6--8\%; we adopt that method-level uncertainty scale here rather than assigning an unsupported percent-level precision to the 0.496~pF value. This range is inherited from the method-level calibration estimate in Ref.~\cite{Panamaldeniya2026_ICPC}; the archived SAP22 analysis does not contain a separate component-by-component uncertainty budget that would justify a more granular uncertainty assignment.

The detector capacitance was independently calculated using the electrostatic model described in Section~\ref{sec:simulation}, yielding 0.488~pF at 700~V. The nominal experiment--simulation difference is 1.6\%, which is smaller than the estimated 6--8\% experimental uncertainty and should therefore be interpreted as consistency within the measurement precision, not as a 1.6\%-level validation of the model.

The relative C--V measurement and the absolute capacitance extraction serve different purposes in this work. The relative C--V curve is used to identify the onset of the depletion plateau, whereas the calibrated charge-injection calculation provides the absolute detector capacitance. The experimentally extracted and simulated capacitance values are compared with those of the previously reported winged SAP17 detector in Table~\ref{tab:SAP17_SAP22_comparison}.

\section{Electrical characterization and source-correlated response}
\label{sec:Electrical characterization}

Radiation interacting with the detector generates electron--hole pairs,
and can produce an additional source-correlated current above the intrinsic leakage current. For an energy deposition $E$ in Ge, the average
number of electron--hole pairs is

\begin{equation}
N_{eh}=\frac{E}{\varepsilon_{\mathrm{Ge}}},
\end{equation}

where $\varepsilon_{\mathrm{Ge}}$ is the average energy required to
create one electron--hole pair in Ge~\cite{Knoll2010_RDM}. At cryogenic
temperature, $\varepsilon_{\mathrm{Ge}}\approx2.96~\mathrm{eV}$; thus,
a fully absorbed 59.5-keV photon from $^{241}$Am generates approximately
$2.0\times10^{4}$ electron--hole pairs.

To distinguish what is directly measured from its physical interpretation, the \emph{leakage current}, $I_{\mathrm{dark}}$, is defined as the current measured in the absence of a radioactive source. The directly observed source-correlated excess is defined as

\begin{equation}
\Delta I_{\mathrm{src}}=I_{\mathrm{meas}}-I_{\mathrm{dark}},
\label{eq:current_components}
\end{equation}

where $I_{\mathrm{meas}}$ is the total current with the source present. Radiation-generated electron--hole pairs are expected to contribute to $\Delta I_{\mathrm{src}}$, but source-correlated surface/interface effects cannot be excluded by the present measurement. The radiation-generated component expected from carrier creation and collection can be estimated at the order-of-magnitude level as

\begin{equation}
I_{\mathrm{rad}}
\simeq q A f_{\Omega} P_{\mathrm{int}}
\frac{E_{\mathrm{dep}}}{\varepsilon_{\mathrm{Ge}}}
\eta_{\mathrm{cc}},
\label{eq:irad_estimate}
\end{equation}

where $q$ is the elementary charge, $A$ is the source activity,
$f_{\Omega}=\Omega/(4\pi)$ is the geometric solid-angle fraction,
$P_{\mathrm{int}}$ is the probability that the emitted radiation
deposits energy in the active Ge volume, $E_{\mathrm{dep}}$ is the mean
energy deposited per interaction, and $\eta_{\mathrm{cc}}$ is the
charge-collection efficiency. A quantitative evaluation of
Eq.~\ref{eq:irad_estimate} requires the calibrated source activity,
source--detector geometry, aperture dimensions, mean deposited energy,
and charge-collection efficiency. These parameters were not
independently characterized in the present measurement; therefore,
Eq.~\ref{eq:irad_estimate} is used only to describe the expected scaling
of a radiation-generated component. The measured quantity reported below
is $\Delta I_{\mathrm{src}}$, and the data are not used to claim that all
of the excess current arises uniquely from bulk radiation-generated charge.

\begin{figure}
\centering
\includegraphics[width=1\linewidth]{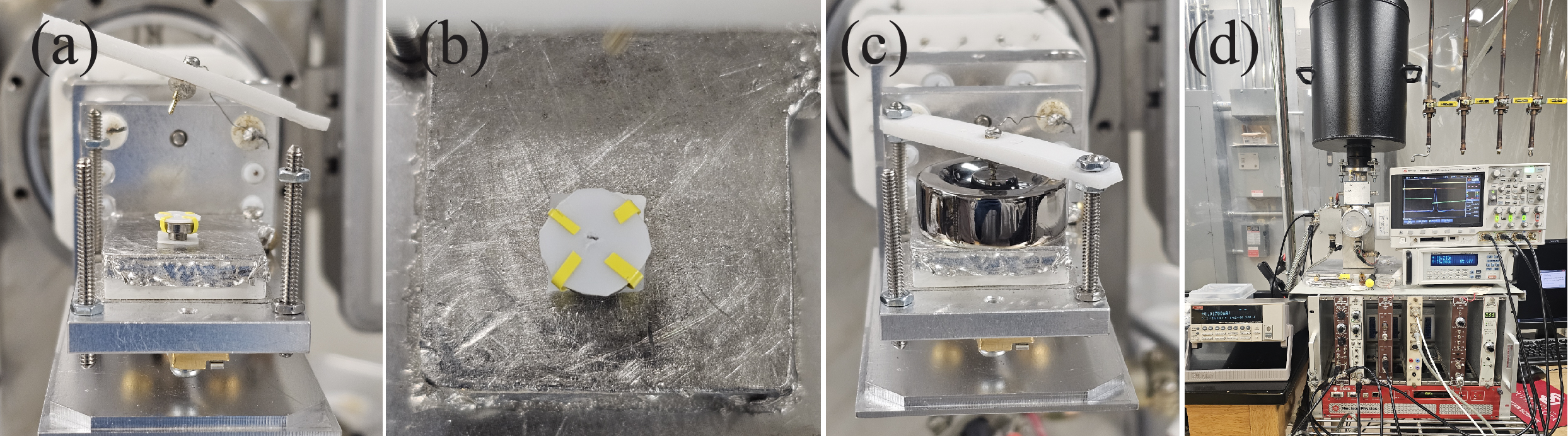}
\caption{Photographs illustrating the $^{241}$Am measurement
configuration. (a) Side view of the $^{241}$Am source assembly mounted
on a PTFE support. (b) Top view of the source showing the PTFE holder
and aperture. (c) Experimental arrangement with the $^{241}$Am source
positioned inside the ICPC bore for localized irradiation.
(d) Complete cryogenic detector setup and associated readout electronics
used for electrical and spectroscopic measurements.}
\label{Alphasetup}
\end{figure}

% OPTIONAL FIGURE ENHANCEMENT: if the original point-by-point current arrays are recovered, add an inset of $\Delta I_{\mathrm{src}}(V)$. The present figure and caption make the measured excess explicit without reconstructing data from the plotted image.
\begin{figure}
\centering
\includegraphics[width=1\linewidth]{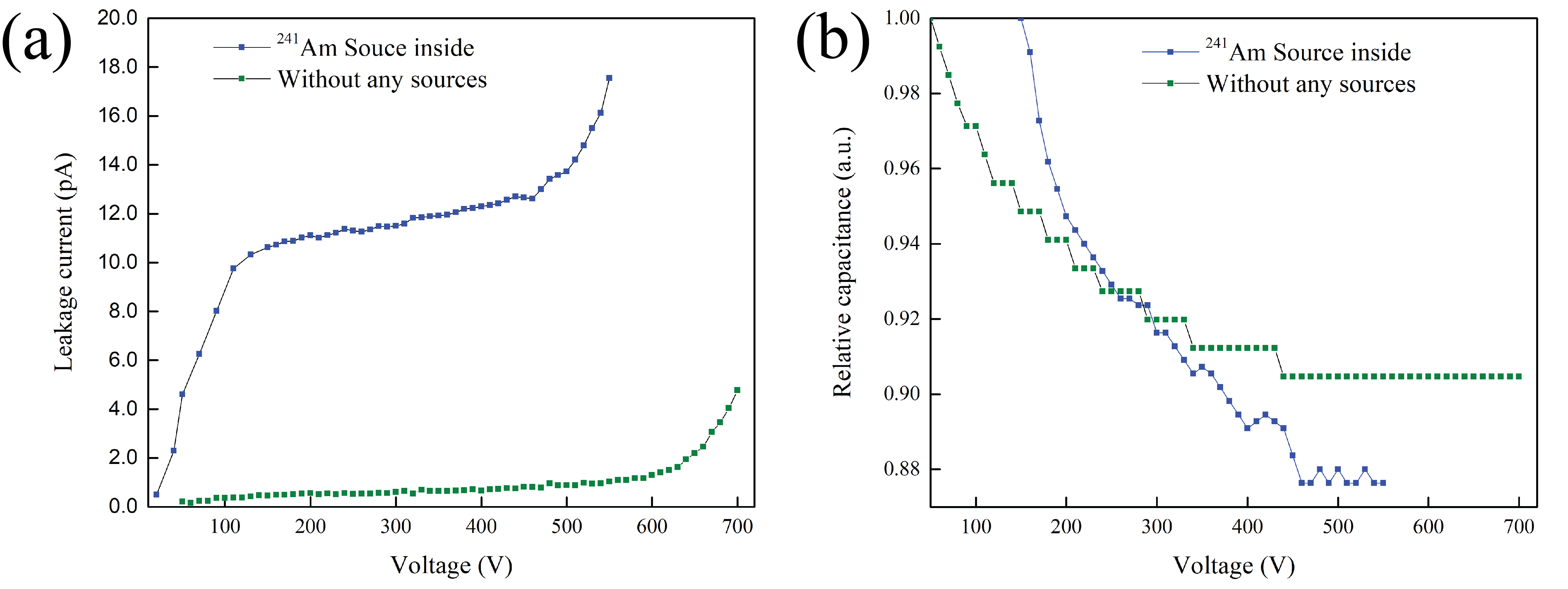}
\caption{(a) Measured current as a function of detector bias and
(b) relative capacitance as a function of detector bias for SAP22,
with and without the $^{241}$Am source positioned inside the detector
bore. The source-free current represents the dark leakage current,
whereas the measurement with the source present contains the dark current
plus a source-correlated excess contribution. At matched bias values, the vertical separation of the two current curves corresponds to $\Delta I_{\mathrm{src}}$ as defined in Eq.~\ref{eq:current_components}. The two relative C--V scans are normalized independently to their respective lowest-bias values and therefore should not be compared as absolute capacitances.}
\label{IV}
\end{figure}

Figure~\ref{IV} compares the measured current and relative capacitance of SAP22 with and without the $^{241}$Am source positioned inside the detector bore. In the absence of the radioactive source, the measured current represents the leakage current and remains in the picoampere regime over most of the investigated bias range. The source-free relative capacitance approaches a plateau near 440~V,
which is taken as the operational experimental indicator of full depletion rather than as a sharply determined depletion threshold.

With the $^{241}$Am source positioned inside the bore, the total measured current increases significantly over the investigated bias range. We report this increase as the source-correlated excess current $\Delta I_{\mathrm{src}}$ defined in Eq.~\ref{eq:current_components}. Radiation-generated electron--hole pairs provide a natural contribution to this excess, but source-induced surface charging, interface effects, or other source-correlated changes in the near-surface region cannot be separated with the present data. The relative C--V response under irradiation exhibits substantial fluctuations and does not show a
well-defined plateau. Consequently, the irradiated relative C--V trace was not used to determine either the depletion voltage or the absolute detector capacitance.

To reduce the radiation flux incident on the detector, the $^{241}$Am source was covered with a PTFE sheet containing a small aperture, as shown in Figure~\ref{Alphasetup}(a) and (b). Measurements performed without this flux reduction produced total measured currents of approximately 40~pA even at low bias. With the PTFE aperture in place, the detector could be biased while still exhibiting a measurable
source-correlated excess current.

The source-correlated excess current under localized $^{241}$Am exposure demonstrates a measurable electrical response to the irradiation, but the excess current by itself does not establish that the contributing interactions are exclusively near the detector surface. The stronger evidence for near-surface sensitivity is the simultaneous observation of the degraded-alpha continuum discussed below. The contact structure consists of approximately 600~nm of amorphous Ge and 120~nm of Al, so alpha particles that reach the detector surface can contribute to the measured response. However, the substantial degradation of the observed alpha energy cannot be attributed to the a-Ge and Al contact layers alone. Energy loss in the source structure, intervening materials, possible surface layers, and incomplete near-surface charge collection may also contribute to the observed degraded-alpha continuum.

This behavior differs from conventional HPGe detectors employing Li-diffused contacts, for which the substantially thicker dead layer can prevent alpha particles from reaching the active detector
volume~\cite{Andreotti2014,Gilmore1995}. The thin a-Ge contact configuration therefore provides sensitivity to near-surface alpha-particle interactions, although quantitative interpretation of
the measured alpha-energy distribution requires additional modeling of the source, intervening materials, contact layers, and near-surface charge collection. A defensible SRIM/Geant4 energy-loss calculation cannot be performed from the archived information because the source-window/encapsulation and PTFE-intervening thicknesses were not independently measured. Accordingly, the present paper does not use the 3558-keV centroid to infer a dead-layer thickness or a unique contact-loss mechanism.

\section{Gamma-ray spectroscopy performance}
\label{sec:Gamma-ray spectroscopy performance}

\subsection{Representative energy spectra}

\begin{figure}
    \centering
    \includegraphics[width=1\linewidth]{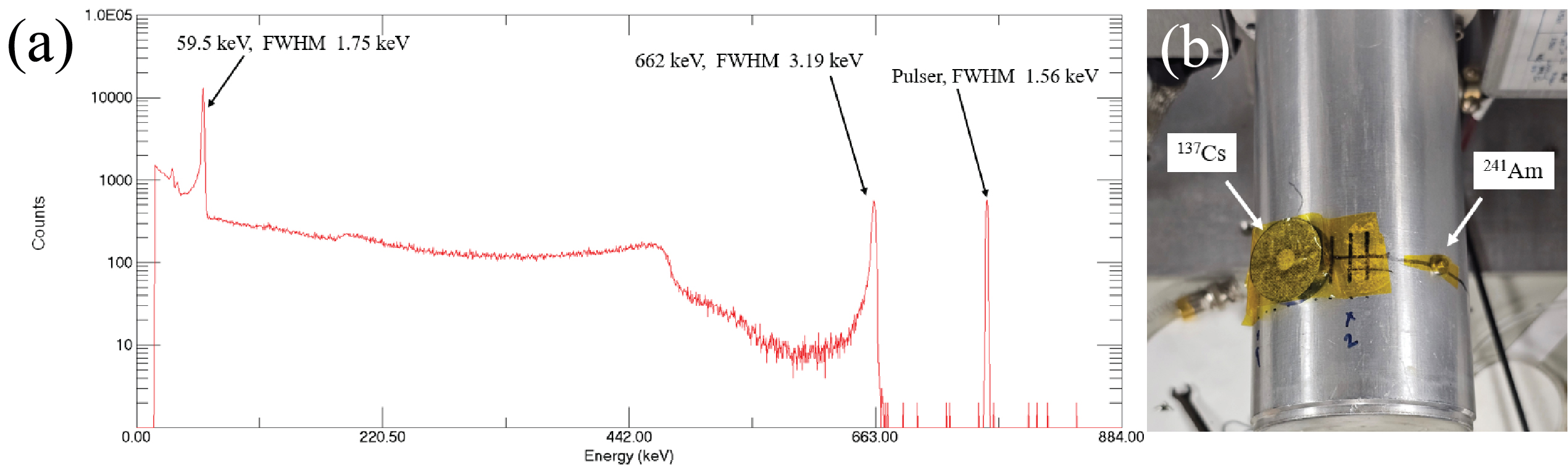}
    \caption{(a) Energy spectrum acquired with the $^{137}$Cs and
    $^{241}$Am sources positioned outside the cryostat at a detector
    bias of 700~V, with the sources at a nominal distance of approximately
    3.5~cm from the detector center. The spectrum had a real time of
    15,017.06~s and a live time of 14,999.90~s. The 59.5-keV $^{241}$Am
    and 662-keV $^{137}$Cs photopeaks are visible together with the pulser peak used to
    monitor the electronic-noise contribution. (b) Photograph of the
    corresponding external-source configuration.}
    \label{both outside}
\end{figure}

\begin{figure}
    \centering
    \includegraphics[width=1\linewidth]{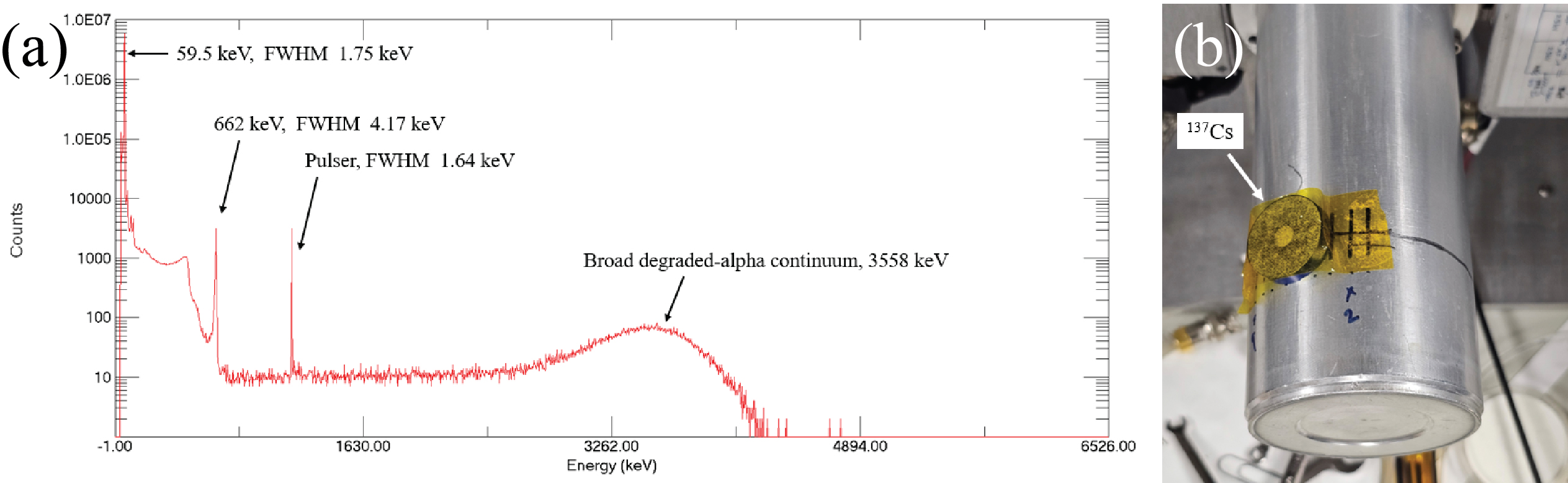}
    \caption{(a) Energy spectrum acquired at a detector bias of
    550~V with the $^{137}$Cs source outside the cryostat and the
    $^{241}$Am source positioned inside the ICPC bore, approximately
    1~mm from the nearest Ge surface. The spectrum had a real time of
    14,955.56~s and a live time of 14,760.18~s. The broad continuum centered near 3558~keV is displayed on
    the gamma-ray-calibrated electron-equivalent energy scale and is
    interpreted as a degraded-alpha response rather than the full
    deposited alpha-particle energy. (b) Photograph of the
    corresponding source configuration.}
    \label{Am241 inside}
\end{figure}

Gamma-ray spectroscopic performance of the SAP22 detector was evaluated using $^{241}$Am and $^{137}$Cs sources at an operating temperature of
77~K. The energy scale was calibrated using the known 59.5-keV photopeak from $^{241}$Am and the 662-keV photopeak from $^{137}$Cs. The energy spectra presented in Figures~\ref{both outside} and ~\ref{Am241 inside} were acquired using a shaping time of 1~$\mu$s. The pulser peak was used to monitor the electronic-noise contribution and to determine the corresponding pulser FWHM on the gamma-ray-calibrated energy scale; it was not used as a substitute for
the gamma-ray energy calibration.

Spectra were acquired under two source configurations. In the external-source configuration, both the $^{241}$Am and $^{137}$Cs sources were positioned outside the cryostat at a nominal distance of approximately 3.5~cm from the detector center, and the detector was operated at a bias of 700~V. Because the detector has finite dimensions, the 3.5~cm value represents the nominal source-placement distance rather than an exact source-to-active-volume distance. The external-source spectrum was acquired for a real time of 15,017.06~s (4.17~h), corresponding to a live time of 14,999.90~s (4.17~h).

In the internal-source configuration, the $^{241}$Am source was positioned inside the ICPC bore region while the $^{137}$Cs source remained outside the cryostat. The $^{241}$Am source surface was approximately 1~mm from the nearest Ge surface; because this spacing was not measured precisely, the value is reported as an estimated source-to-surface separation. The detector was operated at 550~V. The internal-source spectrum was acquired for a real time of 14,955.56~s (4.15~h), corresponding to a live time of 14,760.18~s (4.10~h). The different detector biases and source geometries used for these two configurations are taken into account when interpreting differences in their spectroscopic performance.

Representative spectra are shown in
Figures~\ref{both outside} and~\ref{Am241 inside}, while the corresponding energy-resolution components are summarized in  Table~\ref{tab:resolution}. With both sources outside the cryostat, well-defined photopeaks were observed at 59.5 and 662~keV. At an
operating bias of 700~V, SAP22 achieved energy resolutions of 1.75~keV FWHM at 59.5~keV and 3.19~keV FWHM at 662~keV, demonstrating stable prototype-level spectroscopic performance. These fit values are retained in Table~\ref{tab:resolution}.

When the $^{241}$Am source was positioned inside the ICPC bore, the 59.5-keV photopeak remained clearly observable and a broad degraded-alpha continuum centered near 3558~keV on the gamma-ray-calibrated electron-equivalent energy scale was observed. Broad degraded-alpha responses have previously been reported for thin-contact HPGe detectors employing comparable a-Ge contact structures~\cite{Mei2022}, consistent with the sensitivity of these contacts to near-surface interactions.

The 662-keV photopeak from the externally positioned $^{137}$Cs source broadened from 3.19~keV FWHM in the external-source configuration to 4.17~keV FWHM when the $^{241}$Am source was positioned inside the
bore; the corresponding fit values are 3.19 and 4.17~keV. However, these measurements were performed at different detector biases, 700 and 550~V, respectively. The observed degradation therefore cannot be attributed uniquely to the source-correlated excess current. Instead, it may result from a combination of the reduced detector bias,
source-correlated excess current, changes in near-surface charge collection, and count-rate-related effects. Despite these differences, the 59.5-keV photopeak retained a reported FWHM of 1.75~keV in both configurations (the value retained in the archived fit analysis), and the detector remained spectroscopically operational under localized $^{241}$Am irradiation. The FWHM values are reported to the precision retained in the archived fit analysis for consistency with the pulser widths. Formal covariance-based uncertainties from the original photopeak fits were not retained; therefore, the final digits should not be interpreted as formal statistical uncertainties. A future re-fit of the original MCA spectra would be required to attach rigorous peak-fit uncertainties.

The observation of degraded-alpha events is particularly relevant to the thin-contact configuration. Conventional HPGe detectors employing
thick Li-diffused contacts possess a substantial inactive surface region that can prevent alpha particles from reaching the active germanium volume~\cite{Knoll2010_RDM,Amman2018_aGe}. In contrast, the thin a-Ge contact structure employed in SAP22 permits near-surface alpha-particle interactions to contribute to the measured detector response.

The broad continuum near 3558~keV should not, however, be interpreted as evidence that the energy degradation is produced primarily by the 600-nm a-Ge and 120-nm Al contact layers. The measured alpha response can be influenced by energy loss in the $^{241}$Am source encapsulation or window, the PTFE aperture and other intervening materials, the a-Ge and Al contact layers, possible surface oxide or
contamination, and incomplete charge collection in the near-surface Ge region. Consequently, the structure observed near 3558~keV is interpreted conservatively in the present work as a degraded-alpha
continuum. A quantitative determination of the individual contributions would require a dedicated alpha-transport and near-surface charge-collection model using measured source-window, aperture, and spacing parameters. In the absence of those inputs, the continuum is used only as evidence that near-surface alpha interactions contribute to the detector response, not as a measurement of the inactive-layer thickness.

The measured gamma-ray photopeak width contains contributions from both electronic noise and detector-related broadening. To estimate the non-electronic contribution, the pulser width was subtracted from the
measured gamma-ray photopeak width in quadrature according to

\begin{equation}
\mathrm{FWHM}_{\mathrm{res}}
=
\sqrt{
\mathrm{FWHM}_{\gamma}^{2}
-
\mathrm{FWHM}_{\mathrm{pulser}}^{2}
}.
\label{eq:fwhmres}
\end{equation}

Here, $\mathrm{FWHM}_{\gamma}$ is the measured gamma-ray photopeak
width and $\mathrm{FWHM}_{\mathrm{pulser}}$ represents the electronic
noise contribution. The quadrature subtraction assumes that the electronic-noise and detector-related contributions can be treated as approximately independent Gaussian broadening terms. The resulting
$\mathrm{FWHM}_{\mathrm{res}}$ therefore represents the residual
non-electronic broadening, including contributions from charge
generation and charge-collection variations, and should not be
interpreted as a purely Fano-limited intrinsic resolution.

% REVISION NOTE: If the original MCA peak-fit covariance outputs are recovered, add formal FWHM uncertainties to this table before final submission.
\begin{table}[htbp]
\centering
\caption{Energy-resolution components of SAP22 for the two source
configurations. The $^{137}$Cs source remained outside the cryostat,
while the $^{241}$Am source was positioned either outside the cryostat
or inside the ICPC bore. Values are reported to the precision retained in
the archived fit analysis for consistency with the pulser widths; formal
covariance-based peak-fit uncertainties were not available and are therefore
not assigned here.}
\label{tab:resolution}
\small
\begin{tabular}{lcccccc}
\hline
$^{241}$Am location &
Energy &
Bias &
FWHM$_{\mathrm{pulser}}$ &
FWHM$_{\gamma}$ &
FWHM$_{\mathrm{res}}$ &
Resolution \\
&
(keV) &
(V) &
(keV) &
(keV) &
(keV) &
(\%) \\
\hline

Outside cryostat
& 59.5 & 700 & 1.56 & 1.75 & 0.79 & 2.94 \\

&
662 & 700 & 1.56 & 3.19 & 2.78 & 0.48 \\

Inside ICPC bore
& 59.5 & 550 & 1.64 & 1.75 & 0.61 & 2.94 \\

&
662 & 550 & 1.64 & 4.17 & 3.83 & 0.63 \\

\hline
\end{tabular}
\end{table}

\section{Electrostatic simulation and interpretation}
\label{sec:simulation}

\subsection{Electrostatic simulation framework}

Electrostatic simulations~\cite{Abt2021_SSD} were performed to evaluate the electric-potential and electric-field distributions of the wingless SAP22 ICPC detector and to aid interpretation of the measured electrical and spectroscopic behavior. The detector geometry was constructed using the dimensions summarized in Table~\ref{tab:geometry},
including the central bore, point-contact electrode, and circumferential
groove.

A uniform p-type net impurity concentration of
$N_A-N_D \approx 3\times10^{10}~\mathrm{cm^{-3}}$ was used in the simulation. This value is representative of the detector-grade HPGe material used for fabrication and is consistent with the impurity level
obtained from previous C--V characterization of material from the same crystal-production process. Because the present SAP22 relative C--V measurement was used primarily to identify the onset of the depletion plateau rather than to extract a spatially resolved impurity profile, the impurity concentration was treated as uniform in the electrostatic model. The value $3\times10^{10}~\mathrm{cm^{-3}}$ is therefore an assumed representative input, not a fit to a measured SAP22 impurity profile.

The outer contact was biased at the operating voltage while the point contact was maintained at ground potential. All exposed detector surfaces were treated as electrically passivated boundaries. This represents an idealized boundary condition and does not explicitly include surface-charge distributions, interface states, or surface-leakage conduction. Consequently, the simulation is used to identify the principal bulk electric-field and potential distributions and to support qualitative interpretation of charge transport, rather than to provide a microscopic model of the a-Ge/Ge interface or passivated detector surfaces.

% REVISION NOTE: A future impurity-sensitivity scan (e.g. 2, 3, 4 x 10^10 cm^-3) would strengthen the depletion-voltage comparison; no such unarchived result is invented here.
\subsection{Electric field and potential distributions}

\begin{figure}
    \centering
    \includegraphics[width=1\linewidth]{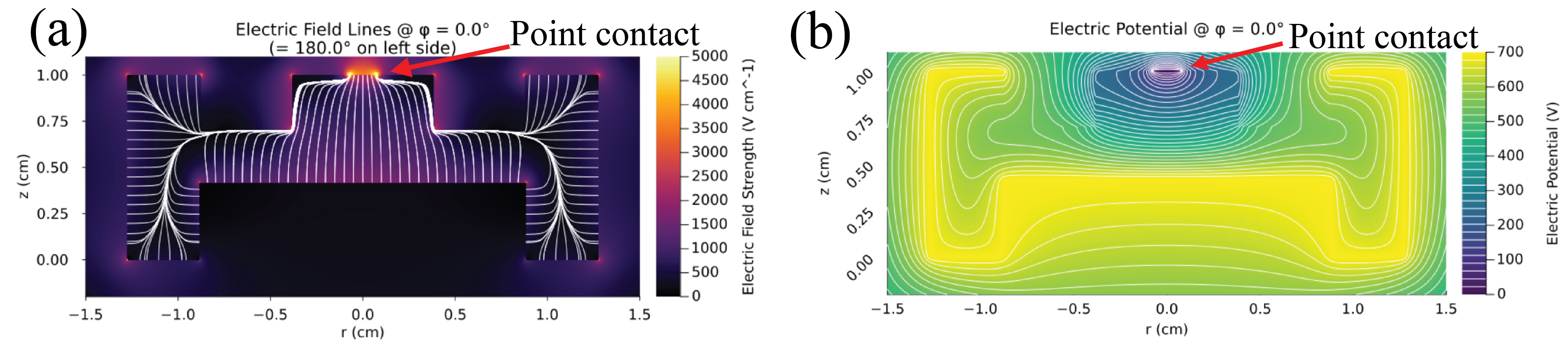}
\caption{Electrostatic simulation of the wingless ICPC detector at a bias voltage of 700~V. (a) Electric-field distribution and field lines in the detector cross section. (b) Electric-potential distribution with equipotential contours. The point-contact electrode is indicated in both panels.}

    \label{fig:field_potential}
\end{figure}

Figure~\ref{fig:field_potential} shows the simulated electric-field and electric-potential distributions of the wingless ICPC detector. As expected for an ICPC geometry, the electric field is strongly concentrated near the point-contact electrode, where the electric-potential gradient is largest. Peak field strengths approach \(5\times10^{3}\,\mathrm{V\,cm^{-1}}\), while the majority of the detector volume remains at substantially lower field strength.

For the bias polarity used here, holes generated in the p-type detector drift toward the grounded point contact, while electrons drift toward the positively biased outer electrode. The strong localized electric field near the point contact promotes rapid final hole collection. In point-contact detectors, signal formation is especially sensitive to carrier motion near the small readout electrode because the weighting potential is strongly localized in that region~\cite{Cooper2011_PSA}. The corresponding electrostatic-potential distribution exhibits closely spaced equipotential contours near the point contact and more widely spaced contours throughout the detector bulk, reflecting the transition from the localized high-field collection region to the lower-field bulk drift region.

This field configuration is consistent with the measured low capacitance and stable spectroscopic performance of the detector. Furthermore, the presence of weak-field regions near portions of the detector surface may influence charge transport for near-surface interactions. Such effects are particularly relevant for the degraded-alpha measurements presented in this work, where alpha particles deposit energy close to the detector surface and may experience partial charge collection before reaching the high-field collection region.

% FINAL FIGURE NOTE: if Figure10.jpg is regenerated from the simulation source, enlarge internal axis/colorbar labels and the point-contact annotation for JINST print readability.

Capacitance was also simulated using the Julia-based framework \texttt{SolidStateDetectors.jl}~\cite{Abt2021_SSD}. At 700~V, the simulated detector capacitance is 0.488~pF, compared with the experimentally extracted value of 0.496~pF obtained using the calibrated charge-injection method described in Section~3.2. The nominal difference is 1.6\%, but this is smaller than the estimated 6--8\% relative experimental uncertainty of the charge-injection method and is therefore interpreted as consistency rather than percent-level model validation. Figure~\ref{IV}(b), in contrast, shows the relative experimental C--V response used to identify the depletion plateau and is not the source of the 0.496~pF absolute-capacitance value.

The present simulation was not used to fit or independently predict a precise depletion voltage. A quantitative depletion-voltage sensitivity study would require additional \texttt{SolidStateDetectors.jl} runs over a plausible impurity range (for example, $2$--$4\times10^{10}~\mathrm{cm^{-3}}$) and, ideally, a measured spatial impurity profile for SAP22. Because those simulation scans are not part of the archived analysis, the model is used here to interpret the bulk field and potential distributions and to provide a capacitance consistency check. Detailed weighting-potential, charge-drift, surface-charge-collection, and impurity-sensitivity calculations are beyond the scope of the present study.

\section{Discussion}
\label{sec:Discussion}

The leakage-current and capacitance--voltage measurements demonstrated stable detector performance over the investigated bias range. In the absence of radioactive sources, the leakage current remained in the picoampere regime and the relative C--V response approached a plateau near 440~V, which is taken as the operational indicator of full depletion rather than as a sharply determined depletion threshold. These observations are consistent with effective charge blocking and surface passivation provided by the thin a-Ge contacts.

The gamma-ray spectroscopy results are consistent with the electrical characterization. In the external-source configuration, the detector achieved energy resolutions of 1.75~keV FWHM at 59.5~keV and 3.19~keV FWHM at 662~keV at an operating bias of 700~V. At $59.5~\mathrm{keV}$, the measured photopeak width ($1.75~\mathrm{keV}$) is dominated by the electronic-noise contribution, as indicated by the pulser FWHM of $1.56~\mathrm{keV}$, whereas at $662~\mathrm{keV}$ the residual non-electronic broadening becomes the dominant contribution. The low detector capacitance characteristic of the ICPC geometry ($0.496~\mathrm{pF}$ measured and $0.488~\mathrm{pF}$ simulated) is favorable for reducing capacitance-related electronic noise.

These results indicate stable signal formation and are consistent with effective bulk charge collection under the tested source configurations; they do not by themselves establish spatially uniform charge-collection efficiency throughout the full detector volume.

When the $^{241}$Am source was positioned inside the ICPC bore region, a moderate degradation of the 662~keV energy resolution was observed, with the FWHM increasing from 3.19~keV to 4.17~keV, while the 59.5~keV photopeak resolution remained 1.75~keV FWHM. It should be noted that the 662~keV measurement with the source inside the bore was performed at a reduced bias of 550~V compared to 700~V for the external-source configuration. 

The degradation of the 662~keV energy resolution may be associated with the source-correlated excess current and the reduced operating bias used during the internal-source measurements. In contrast, the 59.5~keV photopeak FWHM is 1.75~keV in both configurations; because formal peak-fit uncertainties were not retained in the archived analysis, no stronger statistical statement is made. A notable result is the observation of a broad degraded-alpha continuum centered near 3558~keV when the $^{241}$Am source was placed inside the detector bore.

The observation of the continuum is enabled by the thin-contact configuration, but its centroid cannot be explained quantitatively from the a-Ge/Al thicknesses alone with the information presently available. The observed continuum lies well below the principal $^{241}$Am alpha-particle energies of approximately 5.486 and 5.443~MeV~\cite{Nesaraja2015_A241} and may reflect energy loss in the source encapsulation and intervening materials together with detector-contact losses, geometry, and partial near-surface charge collection. Such measurements are generally inaccessible in conventional lithium-diffused HPGe detectors, whose dead layers are typically much thicker than the range of alpha particles in germanium~\cite{Gilmore1995, Andreotti2014}. Thin a-Ge contacts therefore enable studies of near-surface interactions while maintaining effective charge blocking and low leakage current. For rare-event applications, this enhanced surface sensitivity is both an opportunity and a background-control challenge: alpha and beta interactions that would be attenuated in a thick Li-diffused layer can become detectable. Future deployment of this contact concept in low-background experiments will therefore require careful surface cleanliness and radiopurity, stable passivation, and dedicated surface-event rejection through pulse-shape analysis and/or detector-field optimization.

From a fabrication perspective, eliminating the winged handling structure simplifies detector production and avoids the machining steps used to form the protective wings. This is expected to improve utilization of detector-grade germanium, which is costly and limited in availability, but the present study does not quantify the mass saved because the original machining-mass records are incomplete. The demonstrated result is therefore successful fabrication and handling without wings, rather than a measured material-saving percentage. The data indicate that the wings primarily served as a fabrication aid and are not required for successful detector operation.

\section{Conclusion}
\label{sec:Conclusion}

A wingless p-type ICPC HPGe detector employing thin amorphous-germanium contacts was successfully fabricated and characterized using a USD-grown, zone-refined germanium crystal. The detector exhibited stable electrical performance, with leakage currents remaining in the picoampere regime and an operational depletion voltage of approximately 440~V identified from the onset of the relative C--V plateau.

Gamma-ray spectroscopy measurements yielded energy resolutions of 1.75~keV FWHM at 59.5~keV and 3.19~keV FWHM at 662~keV, showing stable prototype-level spectroscopic performance after removal of the winged support structure; the corresponding fit values are retained in Table~\ref{tab:resolution}. The present resolution is not superior to the best value obtained with the earlier winged prototype, so the principal demonstrated advantage is fabrication without wings rather than improved energy resolution. Under localized irradiation with a $^{241}$Am source positioned inside the ICPC bore region, the detector maintained spectroscopic operation despite a source-correlated excess current.

The thin-contact structure permitted observation of a broad degraded-alpha continuum that is consistent with sensitivity to near-surface alpha interactions. Such interactions are strongly attenuated by the thick Li-diffused outer contacts used in conventional p-type HPGe detectors. Because the source-window/intervening-material parameters and near-surface collection response were not independently measured, the continuum is not used to infer a unique dead-layer thickness or energy-loss mechanism. The result nevertheless highlights the potential of thin a-Ge contacts for near-surface interaction studies and surface-event characterization.

Overall, the results demonstrate stable electrical behavior, sub-pF capacitance, and prototype-level spectroscopic performance in a wingless thin-contact ICPC detector that is substantially larger than SAP17. The successful fabrication and handling of the detector without protective wings supports a simplified fabrication route and eliminates wing-related machining; a fully quantitative material-saving claim will require the original machining-mass records. For rare-event applications, the same thin-contact surface sensitivity that expands access to low-energy and near-surface interactions also increases the importance of radiopure surfaces and surface-event discrimination. With those considerations, the wingless geometry provides a practical platform for further development of low-background ICPC HPGe detectors. SAP22 remains prototype-scale, and demonstrating the same fabrication and electrical advantages at substantially larger detector masses is a logical next step.

\section{Credit authorship contribution statement}

S.~A.~Panamaldeniya: Detector fabrication, data analysis, and manuscript writing.
K.~M.~Dong: Electrostatic simulations and simulation analysis.
D.~Mei: Conceptualization, methodology, supervision, and manuscript writing/revision.

\section*{Acknowledgements}
% The authors would like to thank... (Acknowledgements text goes here.)
This work was supported in part by the U.S. National Science Foundation under Grants No. OISE-1743790, OIA 2437416, and PHYS-2310027, and by the U.S. Department of Energy under Grants No. DE-SC0024519 and DE-SC0004768. This research was also supported by a research center funded by the State of South Dakota. Additional support was provided by the U.S. Air Force Research Laboratory under Award No. FA9550-23-1-0495, titled “Building Artificial Intelligence Research Capacity at the University of South Dakota.” We acknowledge Lawrence Berkeley National Laboratory for providing the cryostat used for detector characterization in this work.

%\appendix
\begin{appendices}

\end{appendices}

\end{document}